\documentclass[preprint,aps]{revtex4}

\usepackage{graphicx}% Include figure files
\usepackage{dcolumn}% Align table columns on decimal point
\usepackage{bm}% bold math
\usepackage{epsf}

\newcommand{\be}{\begin{eqnarray}}

\newcommand{\Eq}[1]{Eq.~(\ref{#1})}
\newcommand{\Eqs}[2]{Eqs.(\ref{#1},\ref{#2})}

\newcommand{\ur}[1]{(\ref{#1})}
\newcommand{\urs}[2]{(\ref{#1},\ref{#2})}
\newcommand{\urss}[3]{(\ref{#1},\ref{#2},\ref{#3})}

\newcommand{\beq}{\begin{equation}}
\newcommand{\eeq}{\end{equation}}

\newcommand{\la}[1]{\label{#1}}
\newcommand{\bea}{\begin{eqnarray}}
\newcommand{\eea}{\end{eqnarray}}
\newcommand{\beqa}{\begin{eqnarray}}
\newcommand{\eeqa}{\end{eqnarray}}
\newcommand{\ba}{\begin{array}}
\newcommand{\ea}{\end{array}}

\newcommand{\half}{{\textstyle{\frac{1}{2}}}}

\newcommand{\n}{\nonumber}

\newcommand{\Tr}{{\rm Tr}}

\newcommand{\eps}{\epsilon^{\kappa\lambda\mu\nu}}
\newcommand{\bydef}{\stackrel{d}{=}}

\def\appendix{\par
\setcounter{subsection}{0}
\setcounter{equation}{0}

\def\thesection{Appendix}
\def\theequation{\Alph{section}.\arabic{equation}}}

\begin{document}

\title{\bf Towards lattice-regularized Quantum Gravity}

\author{Dmitri Diakonov}

\affiliation{
Petersburg Nuclear Physics Institute, Gatchina 188300,
St. Petersburg, Russia}

\date{started: August 12, 2011}
\date{August 31, 2011}

\begin{abstract}

Using the Cartan formulation of General Relativity, we construct a well defined lattice-regularized theory capable to describe large non-perturbative quantum fluctuations of the frame field (or the metric) and of the spin connection. To that end we need to present the tetrad by a composite field built as a bilinear combination of fermion fields. The theory is explicitly invariant under local Lorentz transformations and, in the continuum limit, under general covariant transformations, or diffeomorphisms. Being well defined for large and fast varying fields at the ultraviolet cutoff, the theory simultaneously has chances of reproducing standard General Relativity in the infrared continuum limit. The present regularization of quantum gravity opens new possibilities of its unification with the Standard Model. \\



%(Physics and Astronomy Classification Scheme, http://www.aip.org/pacs/pacs.html/)

\end{abstract}

\maketitle

\section{Introduction}

One of the most profound problems in Quantum Field Theory is to construct a well-defined
quantum gravity. Despite many ingenious attempts in the past, General Relativity remains today the only fundamental field theory where quantum fluctuations are untamed. The classical metrics and curvature follow from the Einstein equation. However we need a theory allowing for quantum fluctuations of metrics about the classical values. This is important not only for a consistent treatment of many gravity-related phenomena but also for a future unification of gravitation with the Standard Model where quantum fluctuations are under control and well understood.

Although we live in a world with Minkowski signature, we wish first of all to define quantum gravity in a world with Euclidian signature. First, we believe that if the partition function of a theory does not exist when Euclidian signature is chosen, its Minkowski counterpart will also be inevitably and incurably sick. It may be concealed in perturbation theory when only small fluctuations are considered but will show up when they are allowed to be arbitrarily large. Second, the Euclidian formulation has its own right, for example in problems related to thermodynamics and to tunneling, like in the Hawking radiation problem where paradoxes are encountered just because we do not know how to quantize Euclidian gravity. Therefore, for clearness we shall discuss here Euclidian gravity. If a theory is well defined for Euclidian signature, it is usually not hard to Wick-rotate it to the Minkowski world.

One of the much discussed difficulties in building quantum gravity is the lack of renormalizability of the standard Einstein theory with matter. In loose terms, the Einstein--Hilbert action does not restrict enough high-momenta fluctuations of the metrics. To overcome the difficulty, $R^2$ and/or supersymmetric modifications of gravity have been suggested. However, there is a far worse difficulty which shows up in $R^2$ or $R^4$ or $R^{100}$ gravity as well and in fact for {\em any} generally covariant action:
Large-amplitude fluctuations of the metrics are not restricted as a matter of principle!

The point is, we live in a world with fermions, and that means that General Relativity must be formulated {\it \`a la} Cartan~\cite{Cartan:1923} when the frame field $e^A_\mu$ (also called tetrad, {\it vierbein} or {\it rep\`ere}) and the spin connection $\omega^{AB}_\mu$ being the gauge field of the Lorentz $SO(4)$ group, are used as 16+24=40 independent field variables, see Section II. In this formulation, the cosmological term (or invariant volume) is
\beq
S_{\rm cosm}=\int\!d^4x\,\det(e)
=\int\!d^4x\,\frac{1}{4!}\,\eps\,\epsilon_{ABCD}\,e^A_\kappa e^B_\lambda e^C_\mu e^D_\nu
\la{cosm}\eeq
and the Einstein--Hilbert--Cartan action is
\beq
S_{\rm EHC}=\int\!d^4x\,\frac{1}{4}\,\eps\,\epsilon_{ABCD}\,
{\cal F}^{AB}_{\kappa\lambda}\,e^C_\mu e^D_\nu
\la{EHC}\eeq
where ${\cal F}_{\mu\nu}^{AB}=[D_\mu D_\nu]^{AB} = \partial_\mu\omega^{AB}_\nu
- \partial_\nu\omega^{AB}_\mu + \omega^{AC}_\mu\omega^{CB}_\nu
- \omega^{AC}_\nu\omega^{CB}_\mu$ is the Cartan curvature and $D_\mu^{AB}=(\partial_\mu+\omega_\mu)^{AB}$ is the covariant derivative.
At the saddle point in $\omega_\mu$, corresponding to zero torsion, this term reduces
to the Einstein--Hilbert action $R\det(e)$ if one assumes the standard relation with the metric tensor, $g_{\mu\nu}=e^A_\mu e^A_\nu$.

According to the general lore of Quantum Field Theory, the exponent of the action
gives the weight for quantum fluctuations of the fields. Both the above actions are not sign-definite and therefore cannot restrict path integrals over the $e_\mu$ and $\omega_\mu$ fields. Indeed, if the frame field is allowed to fluctuate, as supposed in quantum gravity, the sign of $\det(e)$ can continuously change from positive to negative or {\it vice versa}. Of course, $\det(e)=0$ is a singular point where the curved space effectively looses one dimension but it is not possible to forbid such local happenings in the world with a fluctuating metric. The Einstein--Hilbert action is double vulnerable: It can change sign both from the flip of the orientation of the frame and from the flip of the curvature sign.

In the next orders, one can generally build 10 invariants quadratic in the Cartan curvature and 5 invariants quadratic in torsion~\cite{Diakonov:2011fs,Baekler:2011jt}. However, all those invariants and in fact all thinkable local actions invariant under diffeomorphisms are necessarily linear in the antisymmetric tensor $\eps$ and hence are not sign-definite!
We discuss it in more detail in Section III.

Therefore, any general-covariant action cannot restrict large fluctuations of $e_\mu$ and $\omega_\mu$, even though small fluctuations about a particular metric may be locally meta-stable. Whatever is the sign with which we take a particular action term in the exponent to define a quantum path integral, there will always be a direction in the functional space where fluctuations are exponentially enhanced instead of being suppressed. We call it the Sign Problem of quantum gravity and it is far worse than the lack of renormalizability. The latter may be cured, at least in perturbation theory, by taking higher derivative terms, but it does not help at all to solve the former.

At this time, we see only one way to overcome the Sign Problem, and that is to use in part
fermionic variables in formulating General Relativity, rather than only bosonic ones. Integrals over anticommuting Grassmann variables are well defined irrespectively of the overall sign in the exponent of a fermionic expression. The reason is that in fermionic integrals introduced by Berezin~\cite{Berezin:1966} one picks up only certain finite order in the Taylor expansion of the exponent of the action, such that the overall sign does not matter.

More specifically, we suggest to present the frame field $e_\mu$ as a composite field bilinear in the anticommuting fermion operators $\psi^\dagger,\psi$ (Section IV). Such kind of expressions has been put forward previously~\cite{Akama:1978pg,Volovik:1990,Wetterich:2005yi,Wetterich:2011yf}, however,
taken literally, our presentation for $e_\mu$ is new. We want to preserve the local gauge Lorentz symmetry exactly at all stages, and for that we need the explicit gauge field $\omega_\mu$. It is only the frame field $e_\mu$ that we replace by a composite fermionic combination, and we use the covariant derivative there with a generic gauge field $\omega_\mu$. Instead of Cartan's $e_\mu$ and $\omega_\mu$, our basic independent variables are $\psi^\dagger,\psi$ and $\omega_\mu$. In this paper, $\psi^\dagger,\psi$ can be thought of as abstract anticommuting variables, however for the goal of unification with the Standard Model they may be in future replaced by real matter fields.

With the fermionic presentation of the tetrad, the Sign Problem above is solved since large-amplitude fluctuations of the metric become restricted. More precisely, their contribution to the partition function becomes finite. Therefore, one can now think of taming also the high-momenta or the short-distance fluctuations by imposing the ultraviolet regularization of the theory. Lattice discretization of space is one of the most
clear and straightforward ways to achieve it. Quite recently a lattice regularization using composite frame field similar to but distinct from our has been suggested by
Wetterich~\cite{Wetterich:2011yf}.

We shall regularize the actions \ur{cosm} and \ur{EHC} and, in principle, any other, higher-derivative action by imposing a 4-dimensional hypercubic grid on the internal curved space. Figuratively, it is like drawing a rectangular pattern on a Scottish plaid. The plaid can be arbitrarily curved but the `physics'  should not depend on the way the pattern is drawn.

It is always desirable to perform the regularization in such a way that all classic symmetries of a field theory are preserved. In our case these are the local symmetries of the action terms \urs{cosm}{EHC}, namely
\begin{itemize}
\item (i) invariance under local Lorentz $SO(4)$ transformations,
\item (ii) invariance under general coordinate transformations, or diffeomorphisms.
\end{itemize}
It is not uncommon, however, that when one attempts to regularize a theory, certain classic symmetries have to be sacrificed in favor of others, with the hope that they will be restored in the continuum limit. Sometimes they are indeed restored, sometimes they are not; in the latter case we say that there is a quantum anomaly.

The gauge Lorentz symmetry will be exact in our lattice regularization but the diffeomorphism-invariance is, strictly speaking, achieved only in the continuum limit, therefore it can in principle turn out to gain a quantum anomaly.

In this paper we formulate a quantum version of General Relativity by means of a well-defined and regularized path integral, such that quantum fluctuations of the metrics and of the spin connection are fully under control and tractable (Section V). The big question is whether the theory that is well defined in the ultraviolet possesses a smooth continuum limit, and if its infrared limit coincides with Einstein's General Relativity. This question will be addressed in Section VI. There are good reasons to believe in the positive answer but further work is needed to establish it.

Finally, we make a provocative remark in Section VII, that the spinor fields used to
define the composite frame field may be in fact real matter fermions. With the quantum
fluctuations of metrics now under control for any number of space-time dimensions, it opens new possibilities to unify quantum gravity with the Standard Model. In particular, by counting the number of degrees of freedom we find that the $SO(16)$ gauge theory possessing diffeomorphism invariance in 16 dimensions is privileged. The $SO(16)$ gauge group contains $SO(4)_{\rm Lorentz}$ and the Standard Model's $SU(3)_{\rm color}\times SU(2)_{\rm weak}\times U(1)_{\rm Y}$ groups as subgroups, and its two 128-dimensional spinor representations fall precisely into four generations of fermions.

\section{Cartan formulation of General Relativity}

We live in a world with fermions, and they must be included into General Relativity.
The standard way one couples Dirac fermions to gravity is via the Fock--Weyl
action~\cite{Fock:1929vt,Weyl:1929fm}
\bea\la{Dirac-0}
S_{\rm Dirac}&=&\int d^4x\,\det(e)\,\frac{1}{2}\left(\psi^\dagger\,e^{A\mu}\,\gamma_A\,
{\cal D}_\mu\psi
-\psi^\dagger\overleftarrow{\cal D}_\mu\,e^{A\,\mu}\,\gamma_A\,\psi\right),\\
\n
&&{\cal D}_\mu=\partial_\mu +\frac{1}{8}\omega_\mu^{BC}[\gamma_B\gamma_C],\qquad
\overleftarrow{\cal D}_\mu =\overleftarrow\partial_\mu -\frac{1}{8}\omega_\mu^{BC}[\gamma_B\gamma_C],
\eea
where $\psi^\dagger,\psi$ are the independent 4-component fermion fields assumed to be world scalars, $\gamma_A$ are the four Dirac matrices, $\omega_\mu^{BC}$ is the gauge field of the local Lorentz group, called spin connection, and $e^{A\,\mu}$ is the contravariant (inverse) frame field,
\beq
\det(e)\,e^{A\,\kappa}=\frac{1}{6}\,\eps\,\epsilon_{ABCD}\,e^B_\lambda e^C_\mu e^D_\nu,
\qquad e^{A\,\kappa}e^A_\lambda=\delta^\kappa_\lambda,\qquad
e^{A\,\kappa}e^B_\kappa=\delta^{AB}.
\la{econtra}\eeq
The metric tensor is the usual
\beq
g_{\mu\nu}=e^A_\mu e^A_\nu.
\la{gmunu}\eeq
To incorporate fermions, one needs, therefore, the gauge field $\omega_\mu$ and the frame
field $e_\mu$, which are {\it a priori} independent variables. Therefore, the bosonic part of the General Relativity action must be also constructed from these fields. We are thus bound to the Einstein--Cartan formulation of General Relativity~\cite{Cartan:1923} even if we wanted to avoid it.

In this formulation, the cosmological and Einstein--Hilbert terms take the form of \Eqs{cosm}{EHC}. If we limit ourselves to just these terms the action is at most quadratic in $\omega_\mu$, therefore the saddle-point integration over $\omega_\mu$ is exact. The saddle-point equation on $\omega_\mu$ is
\beq
(D_\mu e_\nu)^A-(D_\nu e_\mu)^A \bydef 2T^A_{\mu\nu}= 0.
\la{torsion-def}\eeq
The l.h.s. of this equation is, by definition, twice the torsion field $T^A_{\mu\nu}$,
therefore, at the saddle point torsion is zero. The solution of \Eq{torsion-def} is the
well known
\beq
\bar\omega^{AB}_\mu(e)=\frac{1}{2}e^{A\kappa}(\partial_\mu e^B_\kappa
-\partial_\kappa e^B_\mu)
-\frac{1}{2}e^{B\kappa}(\partial_\mu e^A_\kappa-\partial_\kappa e^A_\mu)
-\frac{1}{2}e^{A\kappa}e^{B\lambda}e^C_\mu(\partial_\kappa e^C_\lambda
-\partial_\lambda e^C_\kappa).
\la{omega-bar}\eeq
Being substituted back into \Eq{EHC} it gives the Einstein--Hilbert action $\bar R\det(e)$ where $\bar R$ is the standard scalar curvature built from the standard Christoffel symbol $\bar\Gamma_{\kappa\lambda,\mu}
=\half(\partial_\kappa g_{\lambda\mu}+\partial_\lambda g_{\kappa\mu}
-\partial_\mu g_{\kappa\lambda})$. We mark quantities referring to the zero torsion
limit with a bar.

One can build systematically the series of action terms invariant under (i) and (ii), classifying them in the number of covariant derivatives~\cite{Diakonov:2011fs}. In the 0$^{\rm th}$ order there is only the cosmological term \ur{cosm}. In the 1$^{\rm st}$ order there is the Dirac action \ur{Dirac-0} and three other fermionic terms~\cite{Diakonov:2011fs}. In the 2$^{\rm nd}$ order there is the Einstein--Hilbert--Cartan term \ur{EHC} and the $P,T$ odd term first introduced in Refs.~\cite{Hojman:1980kv,Nelson:1980ph}:
\beq
S_{P,T\,{\rm odd}}
=\int\!d^4x\,\frac{1}{2}\,\eps\,{\cal F}^{AB}_{\kappa\lambda}e^A_\mu e^B_\nu.
\la{Holst}\eeq
Also in the 2$^{\rm nd}$ order in the covariant derivatives there are 5 terms quadratic in torsion \ur{torsion-def}, two of which are $P,T$ odd~\cite{Diakonov:2011fs,Baekler:2011jt}. In the 4$^{\rm th}$ order there are 10 terms quadratic in curvature ${\cal F}$ two of which are full derivatives and three being $P,T$ odd~\cite{Diakonov:2011fs,Baekler:2011jt}. In the zero-torsion limit they reduce to only two well-known independent invariants $\bar R^2\det(e)$ and $\bar R_{\kappa\lambda} \bar R^{\kappa\lambda}\det (e)$, and two full derivatives, where $\bar R_{\kappa\lambda}$ is the standard symmetric Ricci tensor. In the 4$^{\rm th}$ order also two terms of the type $T(\nabla R)$ appear~\cite{Diakonov:2011fs}, and so on.

All these action terms are linear in the antisymmetric $\eps$ and are therefore not sign-definite.

\section{The curse of general covariance}

The Sign Problem of quantum gravity and in fact its main problem is that any diffeomorphism-invariant action is not sign definite, therefore it cannot restrict large fluctuations of the metric. The reason is very general. One constructs world scalars by contracting upper and lower indices, contravariant and covariant tensors. The contravariant tensors can be obtained from covariant ones by the `index rising procedure', {\it e.g.} $A^\mu = g^{\mu\nu}A_\nu.$ In its turn, the contravariant metric tensor, being inverse to the covariant one, can be always expressed through it as
\beq
g^{\mu\nu}=\frac{4\,\epsilon^{\alpha_1\alpha_2\alpha_3\mu}\,
\epsilon^{\beta_1\beta_2\beta_3\nu}\,g_{\alpha_1\beta_1}g_{\alpha_2\beta_2}
g_{\alpha_3\beta_3}}{\epsilon^{\alpha_1\alpha_2\alpha_3\alpha_4}\,
\epsilon^{\beta_1\beta_2\beta_3\beta_4}\,g_{\alpha_1\beta_1}g_{\alpha_2\beta_2}
g_{\alpha_3\beta_3}g_{\alpha_4\beta_4}}.
\la{gcontra}\eeq
Therefore, any world scalar can be always written through covariant tensors only using
an even number of antisymmetric {\it epsilon}'s. In the generally covariant action, however, one integrates world scalars over the 4-volume $\int d^4x$, and this must be independent of the change of coordinates $x^\mu\to x^{'\,\mu}(x)$. In other words, the Jacobian $\partial x/\partial x'$ arising from the change of coordinates in the volume element has to be compensated by the change of coordinates in the integrand. This compensation happens in integrands that are linear in the antisymmetric {\it epsilon} with four Greek indices referring to the curved space, like $\det(e)=(1/4!)\,\eps\,\epsilon_{ABCD}\,e^A_\kappa e^B_\lambda e^C_\mu e^D_\nu\;\;$ or $\;\;\eps\, \epsilon_{ABCD}\,{\cal F}^{AB}_{\kappa\lambda}e^C_\mu e^D_\nu$, {\it etc}.

Indeed, under general coordinate transformations $x^\mu \to x^{'\,\mu}(x),$ expressions
of the type $\eps\,T_{\kappa\lambda\mu\nu}$ where $T$ is a covariant tensor or a combination of such tensors, gets the inverse Jacobian factor $\partial x'/\partial x$ which compensates the change of coordinates in the volume element, such that the action is
diffeomorphism-invariant. This is also true for more complicated algebraic constructions
that have one extra {\it epsilon} in the numerator as compared to the denominator. It is
the only kind of actions allowed by general covariance.

Meanwhile, all actions that are odd in the antisymmetric {\it epsilon} are apparently not
sign-definite. The simplest example is the invariant volume itself, $\int d^4x\,\det(e)$.
If the frame field is allowed to fluctuate, as supposed in quantum gravity, the sign of
$\det(e)$ can continuously change from positive to negative or {\it vice versa}. Of course, $\det(e)=0$ is a singular point where the curved space effectively looses one dimension but it is not possible to forbid such local happenings in the world with a fluctuating metric.
The Einstein--Hilbert--Cartan action can change sign both from the flip of the tetrad orientation and from the flip of the curvature sign.

Therefore, none of the general coordinate actions can protect the system from large quantum fluctuations: there will always be a direction in functional space where fluctuations
are exponentially enhanced instead of being suppressed. A simple visualization of the situation is provided by the $\phi^3$ theory (Fig.~1).

\begin{figure}[htb]
\begin{minipage}[]{.99\textwidth}
\includegraphics[width=0.3\textwidth]{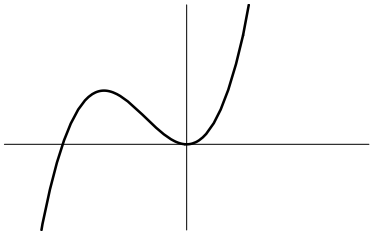}
\end{minipage}
\caption{The $\phi^3$ theory is fundamentally sick both in Euclidean space, where
it is unbounded, and in Minkowski space where it can tunnel to a bottomless state. However, perturbation theory exists in the usual sense near $\phi=0$.}
\label{fig:1}
\end{figure}

It may seem that choosing Minkowski space-time makes the sign problem less acute as the
weight is oscillating anyway independently of the action sign. However the example of the
$\phi^3$ theory shows that it does not necessarily help. In a theory with an unbounded
Lagrangian in both directions either the casual Green function can not be defined, or
there is quantum tunneling to a bottomless state. In any case, non-perturbatively the
theory does not exist.

Since the indefiniteness of the action sign is at the heart of general covariance,
it looks as being the main problem of quantum gravity: the amplitude of quantum fluctuations are not restricted as a matter of principle. This is very different from other fundamental quantum theories.

\section{Composite frame field}

The only way to solve the Sign Problem we see today is to use in part fermionic variables in General Relativity, instead of the bosonic ones. Integration over Grassmann variables
are well defined independently of whether the exponent of the fermionic action is
positive or negative. The reason is that in Grassmann integration one picks up only
a certain finite order in the Taylor expansion of the exponent of the action.

If there are $N$ Grassmann variables $\psi_i$ and $N$ Grassmann variables $\psi^\dagger_i$, such that $\psi_i\psi_j=-\psi_j\psi_i$, $\psi^\dagger_i\psi_j=-\psi_j\psi^\dagger_i$,
$\psi^\dagger_i\psi^\dagger_j=-\psi^\dagger_j\psi^\dagger_i$, one defines, following Berezin~\cite{Berezin:1966},
\beq
\int\prod_{i=1}^N d\psi^\dagger_i\prod_{j=1}^N d\psi_j=0,\qquad
\int\prod_{i=1}^N d\psi^\dagger_i\, \psi^\dagger_i
\prod_{j=1}^N d\psi_j\,\psi_j\;=\;1.
\la{Grassmann}\eeq
Using this definition one obtains
\bea\n
&&\int\prod_{i=1}^Nd\psi^\dagger_i\prod_{j=1}^N d\psi_j\,
\exp\left(\psi^\dagger_i\,A_{ij}\,\psi_j\right)
\quad =\quad \frac{1^2}{N!}\,\epsilon^{i_1...i_N}\epsilon^{j_a...j_N}\,
A_{i_1j_1}\ldots A_{i_Nj_N}\;=\det(A),\\
\n
&&\int\prod_{i=1}^Nd\psi^\dagger_i\prod_{j=1}^N d\psi_j\,
\exp\left(\psi^\dagger_i\psi^\dagger_j\,A_{ij,kl}\,\psi_k\psi_l\right)\\
\n
&& =\frac{2^2}{N!}\,\epsilon^{i_1i_2...i_{N-1}i_N}\epsilon^{j_1j_2...j_{N-1}j_N}\,
A_{i_1i_2,j_1j_2}\ldots A_{i_{N-1}i_N,j_{N-1}j_N}\qquad(N={\rm even}),\quad etc.
\eea
In fact, integrals with any (even) number of fermion variables in the exponent are
well defined and finite, irrespectively of the overall sign in the exponent.

We wish to use this `fermionization' trick to define quantum gravity theories
allowing for arbitrary metric fluctuations, small and large. By a `quantum gravity theory'
we mean a theory with a local action invariant under gauge transformations of the Lorentz group, and under diffeomorphisms. In particular, we shall keep in mind the most common actions \urss{cosm}{EHC}{Dirac-0}.

More specifically, we introduce the frame field $e^A_\mu$ as a composite field
bilinear in anticommuting fermion operators:
\beq
\hat e^A_\mu = \frac{1}{2}\left(\psi^\dagger\gamma_A{\cal D}_\mu\psi
-\psi^\dagger\overleftarrow{\cal D}_\mu\gamma_A\psi\right).
\la{composite}\eeq
This expression transforms as a world vector under diffeomorphisms, and as a vector under
{\em local} Lorentz transformations, which are the correct transformation rules for the frame field, as seen from the Dirac--Fock--Weyl action \ur{Dirac-0}.

The idea of using the composite frame field appears first in a paper by
Akama~\cite{Akama:1978pg}. It reappears independently in Volovik's derivation of ``general
relativity'' description of the $^3{\rm He}$-B superfluid~\cite{Volovik:1990} where
$\psi^\dagger, \psi$ are real matter fields. More recently, a composite frame field
of the type \ur{composite} has been considered by
Wetterich~\cite{Wetterich:2005yi,Wetterich:2011yf}.
In Refs.~\cite{Volovik:1990,Wetterich:2005yi,Wetterich:2011yf} the ordinary (rather than covariant) derivative is used in the definition \ur{composite}, such that $e^A_\mu$ does not transform as it should under {\em local} Lorentz transformations~\cite{footnote-1}.
In Ref.~\cite{Akama:1978pg} the covariant derivative is used in the construction but with
the spin connection taken at its saddle-point value \ur{omega-bar} expressed in its turn
through the tetrad, which makes the definition somewhat ambiguous. Therefore, taken literally, \Eq{composite} is new.

We stress that in the Cartan formulation, the frame field $e^A_\mu$ and the spin
connection $\omega^{AB}_\mu$ are independent variables. In \Eq{composite} only the frame field is replaced by the composite expression through spinor fields whereas the spin connection remains an independent gauge variable. This is not only aesthetically appealing but is also helpful if one keeps in mind possible unification when $\omega_\mu$ is joined by the gauge fields of the Standard Model (Section VII). In the Standard Model we have spinors and gauge fields, why should gravity be different?

The new content of the gravity theory are now the fermion fields $\psi^\dagger, \psi$ and the gauge field of the Lorentz group $\omega^{AB}_\mu$.

\section{Lattice regularization of quantum gravity}

When one substitutes the frame field by its composite expression \ur{composite} the
cosmological term $\det(\hat e)$ becomes quartic in $\psi^\dagger$ and quartic in $\psi$,
whereas the Einstein--Hilbert--Cartan action $({\cal F}\wedge \hat e\wedge \hat e)$ becomes quadratic in $\psi^\dagger$ and quadratic in $\psi$. We denote by the hat the frame
fields replaced by the fermion bilinear expression \ur{composite}.

It is interesting that the Dirac--Fock--Weyl action \ur{Dirac-0} for the same fermions
as in the definition of the frame is nothing but the cosmological term. Indeed, one can
rewrite \Eq{Dirac-0} as
\bea\la{Dirac-1}
S_{\rm Dirac}&=&\int\!d^4x\,\frac{1}{6}\,\eps\,\epsilon_{ABCD}\,\hat e^B_\lambda
\hat e^C_\mu \hat e^D_\nu\,
\frac{1}{2}\left(\psi^\dagger\gamma_A{\cal D}_\kappa\psi -\psi^\dagger
\overleftarrow{\cal D}_\kappa\gamma_A\psi\right) \\
\n
&=&\int\!d^4x\,\frac{1}{6}\,\eps\,\epsilon_{ABCD}\,\hat e^B_\lambda \hat e^C_\mu
\hat e^D_\nu \cdot\hat e^A_\kappa\quad =\quad 4\int\!d^4x\,\det(\hat e)\;=\;4S_{\rm cosm}.
\eea
This relation can be read in the opposite direction: the cosmological term for composite
tetrad is the Dirac--Fock--Weyl action in disguise. In terms of fermions it corresponds
to a propagating theory. Therefore, an interesting possibility opens that when the composite frame field is used, the cosmological term by itself without adding the Einstein--Hilbert action, is capable to reproduce standard classical gravity theory in the infrared limit. We shall discuss this question in the next Section.

In order to give precise sense to the path integrals over the fermion fields $\psi^\dagger,\psi$ and over the spin connection $\omega_\mu$ we discretize the 4-dimensional internal space of Euclidean signature by imposing a hypercubic grid with lattice spacing $a$. We introduce four 4-vectors $a^\nu_\mu = a\delta^\nu_\mu$ to describe
the shifts of lattice sites to four neighbor sites in the positive direction. The coordinates of lattice sites are taken to be integers in units of $a$.

\subsection{Spin connection, link variables}

As in any lattice gauge theory, we replace the connection $\omega_\mu$ by a $4\times 4$ unitary matrix ``living'' on lattice links,
\beq
\Omega_\mu =\exp\left(-a\frac{\omega^{AB}_\mu}{8}[\gamma_A\gamma_B]\right),\quad
\Omega^\dagger_\mu =\exp\left(a\frac{\omega^{AB}_\mu}{8}[\gamma_A\gamma_B]\right),\quad
\Omega^\dagger_\mu \Omega_\mu={\bf 1}_{4\times 4},
\la{link-Omega}\eeq
where $a$ is the lattice spacing. In the spinor basis
\beq
\gamma_A=\left(\begin{array}{cc}0 & \sigma^-_A\\
\sigma^+_A & 0 \end{array}\right),\qquad
[\gamma_A\gamma_B]=2i \left(\begin{array}{cc} \bar\eta^i_{AB}\tau^i & 0\\
0 & \eta^i_{AB}\tau^i  \end{array}\right),
\la{spinor-basis}\eeq
where $\sigma^\pm_A=(\pm i \mbox{\boldmath$\tau$}, 1)$ and $\mbox{\boldmath$\tau$}$
are the three Pauli matrices; $\eta,\bar\eta$ are 't Hooft symbols projecting the general
$so(4)$-valued connection onto its $su(2)_{\rm L}$ and $su(2)_{\rm R}$ parts:
\beq
\omega^{AB}_\mu =-\frac{1}{2}\bar\eta^i_{AB}\,L^i_\mu-\frac{1}{2}\eta^i_{AB}\,R^i_\mu\,.
\la{omega-decomp}\eeq
Correspondingly, link variables are block-diagonal unitary matrices composed of the
$SU(2)_{\rm L}$ and $SU(2)_{\rm R}$ rotations:
\bea\la{Omega-diag}
\Omega_\mu &=&\left(\begin{array}{cc} U_{L\,\mu} & 0 \\ 0 & U_{R\,\mu} \end{array}\right)
=\left(\begin{array}{cc} \exp\left(ia\frac{\tau^a}{2}L^a_\mu\right) & 0 \\
0 & \exp\left(ia\frac{\tau^a}{2}R^a_\mu\right) \end{array}\right),\\
\n\\
\n
\Omega^\dagger_\mu &=&\left(\begin{array}{cc} U^\dagger_{L\,\mu} & 0 \\
0 & U^\dagger_{R\,\mu} \end{array}\right)
=\left(\begin{array}{cc} \exp\left(-ia\frac{\tau^a}{2}L^a_\mu\right) & 0 \\
0 & \exp\left(-ia\frac{\tau^a}{2}R^a_\mu\right) \end{array}\right)\,.
\eea

\subsection{Curvature, plaquette variable}

As usually in lattice gauge theory, the plaquette gives the curvature (see Fig.~2, left):
\bea\la{plaquette-1}
\Omega_{\mu\nu}&=&\Omega_\mu\left(x+\frac{a_\mu}{2}\right)
\Omega_\nu\left(x+a_\mu+\frac{a_\nu}{2}\right)
\Omega^\dagger_\mu\left(x+\frac{a_\mu}{2}+a_\nu\right)
\Omega^\dagger_\nu\left(x+\frac{a_\nu}{2}\right)\\
\n
&=&{\bf 1}_{4\times 4}+\frac{a^2}{8}{\cal F}^{AB}_{\mu\nu}[\gamma_A\gamma_B]
+{\cal O}(a^3),
\eea
where in its turn the $so(4)$-valued curvature can be decomposed into the $su(2)_{\rm L}$
and $su(2)_{\rm R}$ parts,
\beq
{\cal F}^{AB}_{\mu\nu}= -\frac{1}{2}\,F^i_{\mu\nu}(L)\,\bar\eta^i_{AB}-\frac{1}{2}\,F^i_{\mu\nu}(R)\,\eta^i_{AB}.
\la{fsSOdec}\eeq
Here
\bea\n
F^i_{\mu\nu}(L)&=&\partial_\mu L^i_\nu-\partial_\nu L^i_\mu +\epsilon^{ijk}L^j_\mu L^k_\nu,\\
\la{fs}
F^i_{\mu\nu}(R)&=&\partial_\mu R^i_\nu-\partial_\nu R^i_\mu +\epsilon^{ijk}R^j_\mu R^k_\nu
\eea
are the usual Yang--Mills field strengths of the $SU(2)$ Yang--Mills potentials $L^i_\mu$
and $R^i_\mu$.

\begin{figure}[htb]
\begin{minipage}[]{.99\textwidth}
\includegraphics[width=0.27\textwidth]{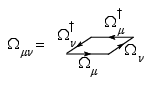}\hspace{1.2cm}
\includegraphics[width=0.25\textwidth]{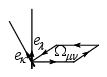}\hspace{1.2cm}
\includegraphics[width=0.25\textwidth]{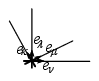}
\end{minipage}
\caption{One plaquette (left); Einstein--Hilbert--Cartan action (middle);
cosmological term action (right).}
\label{fig:2}
\end{figure}

\subsection{Gauge transformation on the lattice}

Under Lorentz gauge transformations ascribed, as usually, to lattice sites, the fermion
field transforms as
\beq
\psi(x)\to V(x)\psi(x) = \left(\begin{array}{cc} V_L & 0 \\ 0 & V_R \end{array}\right)
\left(\begin{array}{c} \psi_L \\  \psi_R \end{array}\right),\qquad
\psi^\dagger(x)\to \psi^\dagger(x)V^\dagger(x),
\la{GT-ferm}\eeq
telling us that the fermion field must `live' on  lattice sites.

The unitary matrix $\Omega_\mu$ ascribed to the link in the $\mu^{\rm th}$ direction
connecting the lattice site with coordinates $x^\nu$ with the lattice site $x^\nu+a^\nu_\mu$ where $a^\nu_\mu=a\delta^\nu_\mu$, transforms as
\bea\la{GT-link}
\Omega_\mu\left(x+\frac{a_\mu}{2}\right) &\to & V(x+a_\mu)\,
\Omega_\mu\left(x+\frac{a_\mu}{2}\right)V^\dagger(x),\\
\n
\Omega^\dagger_\mu\left(x+\frac{a_\mu}{2}\right) &\to & V(x)\,
\Omega^\dagger_\mu\left(x+\frac{a_\mu}{2}\right)V^\dagger(x+a_\mu).
\eea
Finally, the frame field written in the matrix form $e_\mu\bydef e^A_\mu\gamma_A$ transforms as
\beq
e_\mu(x)\to V(x)e_\mu(x) V^\dagger(x).
\la{GT-frame}\eeq
\vspace{-1.5cm}

\subsection{Frame field, lattice site variable}

It follows from the transformation law \ur{GT-frame} that the frame field should be ascribed to lattice sites rather than to links, despite that it carries a ``direction'' index. This is shown symbolically in Fig.~2, middle and right.

There are many discretized expressions for $\hat e^A_\mu$ that transform according to \Eq{GT-frame} and tend to \ur{composite} in the limit $a\!\to\! 0$. For example, one
can define using an arbitrary parameter $\alpha$:
\bea\la{e}
\hat e^A_\kappa(x) &=&
\frac{\alpha}{2a}\,\left[\psi^\dagger(x)\gamma_A
\Omega^\dagger_\kappa \left(x+\frac{a_\kappa}{2}\right)\psi(x+a_\kappa)
- \psi^\dagger(x+a_\kappa)\Omega_\kappa \left(x+\frac{a_\kappa}{2}\right)\gamma_A
\psi(x)\right]\\
\n
&-&\frac{1-\alpha}{2a}\left[\psi^\dagger(x)\gamma_A
\Omega_\kappa \left(x-\frac{a_\kappa}{2}\right)\psi(x-a)
-\psi^\dagger(x-a_\kappa)\Omega^\dagger_\kappa \left(x-\frac{a_\kappa}{2}\right)
\gamma_A\psi(x)\right]\\
\n\\
\n
&\stackrel{a\to 0}{\longrightarrow} & \frac{1}{2}\left(\psi^\dagger\gamma_A{\cal D}_\mu\psi -\psi^\dagger\overleftarrow{\cal D}_\mu\gamma_A\psi\right).
\eea
Keeping in mind that $e^A_\mu$ is a vector in both the inner curved space and in the flat
tangent space and in particular that its components change signs appropriately under $P,T$ inversions in both spaces, it is natural to choose the parameter $\alpha=\frac{1}{2}$
in \Eq{e}. The discretized tetrad is shown schematically in Fig.~3.
\begin{figure}[htb]
\begin{minipage}[]{.89\textwidth}
\includegraphics[width=0.85\textwidth]{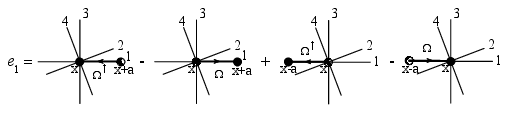}
\end{minipage}
\caption{Four terms for the discretized frame field, \Eq{e}. Solid blobs stand for $\psi^\dagger$, open blobs stand for $\psi$. Lines with arrows denote parallel transporters $\Omega^\dagger, \Omega$.}
\label{fig:3}
\end{figure}

\subsection{Discretized action terms}

In the discretized version, the cosmological term is
\bea\la{cosm-1}
S_{\rm cosm}&=&\int\! d^4x\,\det(\hat e)\\
\n
&\to & \sum_{\rm sites} a^4\,\frac{1}{4\cdot 4!}\,\eps\,\Tr(\hat e_\kappa \hat e_\lambda \hat e_\mu \hat e_\nu\,\gamma_5)
=\sum_{\rm sites} a^4\frac{1}{4!}\,\eps\,\epsilon_{ABCD}\,\hat e^A_\kappa \hat e^B_\lambda \hat e^C_\mu \hat  e^D_\nu\\
\n
&=& \sum_{\rm sites} a^4\,\epsilon_{ABCD}\,\hat e^A_1 \hat e^B_2 \hat e^C_3 \hat e^D_4
\;\equiv\;\sum_{x\,\in\, {\rm sites}} s_{\rm cosm}(x),
\eea
where the summation goes over all lattice sites $x$ and the four frame fields are replaced by their discretized expression \ur{e}. The Einstein--Hilbert--Cartan action is
\bea\la{EHC-1}
S_{\rm EHC} &=& \int\! d^4x\,\frac{1}{4}\eps\,\epsilon_{ABCD}\,
{\cal F}^{AB}_{\kappa\lambda}\hat e^C_\mu \hat e^D_\nu \\
\n
&\to & \sum_{\rm sites} a^2\,\frac{1}{4}\,\eps\,\Tr(\Omega_{\kappa\lambda}
\hat e_\mu \hat e_\nu\,\gamma_5)\;=\;\sum_{\rm sites} a^2\,\frac{1}{4}\,\eps\,\Tr(\Omega_{\kappa\lambda}\gamma_C\gamma_D\gamma_5)
\hat e^C_\mu \hat e^D_\nu\;\equiv\;\sum_{x\,\in \,{\rm sites}} s_{\rm EHC}(x),
\eea
where $\Omega_{\mu\nu}$ is the plaquette variable \ur{plaquette-1}, and the two frame fields are replaced by their discretized expression \ur{e}. Both action terms \ur{cosm-1}
and \ur{EHC-1} are explicitly invariant under local Lorentz gauge transformations and become invariant under diffeomorphisms in the continuum limit. We notice that when the
discretized composite tetrad \ur{e} is used, the lattice spacing factors $a^4,a^2$ drop out from \Eqs{cosm-1}{EHC-1}, respectively.

\subsection{Lattice integration measure}

On the lattice, one integrates over 8 Grassmann variables $\psi^\dagger,\psi$ at each lattice site, keeping in mind that only the following Berezin integral for each lattice site is nonzero:
\beq
\int d\psi^\dagger_1 d\psi^\dagger_2 d\psi^\dagger_3 d\psi^\dagger_4\;
d\psi_1 d\psi_2 d\psi_3 d\psi_4\;\psi^\dagger_1 \psi^\dagger_2 \psi^\dagger_3 \psi^\dagger_4\;\psi_1 \psi_2 \psi_3 \psi_4\;=\;1.
\la{Berezin}\eeq
One integrates also over all link variables $\Omega_\mu$ with the $SO(4)\simeq
SU(2)_{\rm L}\times SU(2)_{\rm R}$ Haar measure normalized to unity. The following
integrals over the $SU(2)$ Haar measure will be used:
\bea\n
&&\int dU_{\rm L,R} = 1,\qquad \int dU\,U^\alpha_\beta
= \int dU\, U^{\dagger\,\alpha}_\beta = 0,\qquad
\int dU\,U^\alpha_\beta U^{\dagger\,\gamma}_\delta =\frac{1}{2}\,\delta^\alpha_\delta\,\delta^\gamma_\beta\,,\\
\la{Haar}
&&
\int dU\,U^\alpha_\beta U^\gamma_\delta=\frac{1}{2}\,\epsilon^{\alpha\gamma}\,\epsilon_{\beta\delta}\,,\qquad
\int dU\,U^{\dagger\,\alpha}_\beta U^{\dagger\,\gamma}_\delta=\frac{1}{2}\,\epsilon^{\alpha\gamma}\,\epsilon_{\beta\delta}\,,
\qquad etc.
\eea

The partition function for the regularized quantum General Relativity is
\bea\la{Z}
{\cal Z}&=&\prod_{\rm sites}\int d\psi^\dagger_1 d\psi^\dagger_2 d\psi^\dagger_3 d\psi^\dagger_4\; d\psi_1 d\psi_2 d\psi_3 d\psi_4\prod_{\rm links}dU_L\,dU_R\\
\n
&&\cdot\exp\left(g_1\sum_{x\in {\rm sites}}s_{\rm cosm}(x)
+g_2\sum_{x\in {\rm sites}}s_{\rm EHC}(x)+\ldots\right)
\eea
where the actions are given by \Eqs{cosm-1}{EHC-1} with dimensionless `coupling constants'
$g_{1,2}$. If needed, discretized versions of higher derivative terms can be added.
This integral is well defined and finite for any finite volume.

\section{How to compute the regularized path integral?}

The arising lattice-regularized path integral for quantum General Relativity, although well defined, is quite unusual. The Einstein--Hilbert--Cartan action density is $s_{\rm EHC}\sim \psi^\dagger \psi^\dagger \psi \psi$ and the cosmological term density is $s_{\rm cosm}\sim \psi^\dagger \psi^\dagger \psi^\dagger \psi^\dagger \psi \psi \psi \psi$. There are no
$\psi^\dagger \psi$ terms in the action, therefore there is no `propagator', only the
multi-fermion `vertices'.

This is a completely new kind of integral, and we have little or no experience in dealing with it. There must be approximate methods but they are not developed. One can hardly expect that there is a direct analog of saddle-point approximation for Berezin integrals
over anticommuting variables. Perhaps, mean field methods have a better chance. For example, one can introduce a mean-field tetrad $<\!e^A_\mu\!>$ and a mean-field connection $<\!\Omega_\mu\!>$, replace all composite tetrads except one by the mean $<\!e^A_\mu\!>$ and replace all links except one by the mean $<\!\Omega_\mu\!>$, and then make it self-consistent, that is calculate the mean values from the definition of the average, hence closing the equations for the mean values. It would be interesting to check if the mean-field values satisfy the discretized version of the no-torsion condition \ur{torsion-def}. Similarly, introducing a mean-field plaquette
$<\!\Omega_{\mu\nu}\!>$, one can check if the mean curvature satisfies the Einstein equation.

With the approximate methods not yet developed, we take a glimpse at the exact calculation
of the partition function from its definition \ur{Z}. The only way to get a nonzero result is to organize Berezin integrals \ur{Berezin} in such a way that precisely 4 fermion operators $\psi^\dagger_1\psi^\dagger_2\psi^\dagger_3\psi^\dagger_4$ and 4 fermion operators $\psi_1\psi_2\psi_3\psi_4$ appear at each lattice site. These operators come from Taylor expanding the exponent of the action. Since the action densities are commuting bosonic operators, one can write $\exp\left(\sum_x s(x)\right)=\prod_x \exp\left(s(x)\right)$ and expand each exponent independently, where $s(x)=g_1s_{\rm cosm}(x)+g_2 s_{\rm EHC}(x)$.

Let us first take for simplicity only the cosmological term in the action
($g_2=0,\,g_1\neq 0$). Then there are exactly two ways how the needed number of fermion operators (4 $\psi^\dagger$'s and 4 $\psi$'s per lattice site) is obtained: one either
expands $\exp\left(g_1 s_{\rm cosm}(x)\right)$ at every lattice site to the linear order
in $s_{\rm cosm}(x)$ or expands this exponent to the second order in every even (or odd)
site, taking the zero order expansion at the alternating sites, see Fig.~4, left.

\begin{figure}[htb]
\begin{minipage}[]{.99\textwidth}
\includegraphics[width=0.36\textwidth]{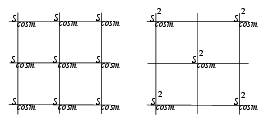}\hspace{0.7cm}
\includegraphics[width=0.57\textwidth]{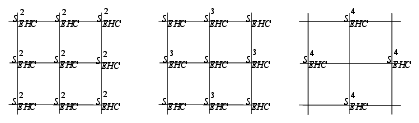}
\end{minipage}
\caption{Two ways of Taylor-expanding the exponent of the cosmological term action (first two figures) and three ways of Taylor-expanding the exponent of the Einstein--Hilbert--Cartan action (last three figures).}
\label{fig:4}
\end{figure}

\noindent
The needed fermion operators at the empty sites are borrowed from the twice filled ones, which is possible since the discretized tetrad \ur{e} has fermion operators belonging to the nearest neighbor site. Similarly, if only the Einstein--Hilbert--Cartan action is used ($g_1=0,\,g_2\neq 0$) there are three patterns of the expansion of the action exponents,
shown in Fig.~4, right. If both action terms are taken into consideration
($g_1\neq 0,\,g_2\neq 0$) it is a combination of the patterns in Fig.~4. We stress that
one never needs to expand the exponents of the action terms beyond the fourth order.

With the exponents of the actions at the lattice sites being expanded, one can perform
explicit integration over the spin connection link variables using \Eq{Haar} and the like.
The result can be only gauge invariant and indeed it is: All fermion operators gather
into gauge-invariant combinations $(\psi^\dagger\psi)$, $(\psi\epsilon\psi)$ and
$(\psi^\dagger\epsilon\psi^\dagger)$ with the fermion operators in the parentheses belonging to the same site. This is a straightforward calculation although the algebra is tedious; it may be trusted to a computer.

Now, one can start integrating over fermion variables keeping in mind Berezin's selection rule \ur{Berezin}. The expression one has to deal with is a product over the volume of
sums of the products of the above gauge-invariant monomials. If at some site one chooses
a particular term in the sum with the necessary number of $\psi^\dagger,\psi$ operators, this term dictates what term should be used at one of the neighbor site, and so on. The rules which terms to use, form close loops on the lattice; there are several kinds of loops. If one goes to another lattice point, the rule for selecting fermionic monomials there can be chosen independently. We arrive, thus, to the presentation of the partition function \ur{Z} as a sum over closed loops of arbitrary length and shape! The details of the derivation will be given elsewhere.

In this language, the question whether the theory possesses a continuum (infrared) limit
is the question whether the portion of long loops in lattice units is sufficiently large. This question is purely combinatorial: one compares the entropy of loops with ``energy'' per unit length but it depends also on the ratio of the constants $g_2/g_1$. This analysis has not been done so far but we see no reasons why the entropy of the loops cannot win over the energy, at least in certain range of $g_2/g_1$. In that case there will be long-range correlations in the system. Since diffeomorphism-invariance is restored in the continuum limit and Lorentz gauge invariance is explicit, these correlations must be equivalent to Einstein's law of gravity, although we have not yet checked it explicitly.

Somewhat surprisingly, the cosmological term or the invariant volume by itself, meaning putting $g_2=0$, may be sufficient to generate the expected classical General Relativity at large distances. As seen from \Eq{Dirac-1}, the cosmological term, when written through the composite frame, is nothing but the Dirac--Fock--Weyl action corresponding to propagating fermions. If some kind of smooth mean field develops in the system, one can imagine integrating out fermions in its background. The low-energy derivative expansion of the effective action for the `emergent gravity' {\it \`a la} Sakharov starts with the Einstein--Hilbert term, therefore the appearance of the standard Generality Relativity for
low momenta mean fields is almost guaranteed. Technically, one has to check if the cosmological term alone in \Eq{Z} is capable of generating long loops, {\it i.e.} long-range correlations.

Apparently \Eq{Z} defines a system with rich dynamics which would be interesting to study more closely.

\section{Possible unification with the Standard Model?}

In \Eq{e} we have replaced the frame field $e^A_\mu$ by a composite expression through spinor fields $\psi^\dagger,\psi$. There is a question of matching the number of degrees of freedom (dof's). If $\psi^\dagger,\psi$ carry less dof's than $e^A_\mu$, not all metric
tensors are possible. If they carry more dof's it is not only pure gravity but something in addition.

In 4 dimensions, $e^A_\mu$ carry $4^2=16$ dof's, whereas one `flavor' of Dirac 4-component spinors $\psi^\dagger,\psi$ carry $4\cdot 2=8$ dof's. From this point of view it may seem preferable to use two `flavors' of fermions to parameterize the tetrad, which of course is also possible. Then the number of dof's in the tetrad and the spinor fields would match each other exactly. However, matching dof's between fermions and bosons is a subtle matter, especially in view of the Nielsen--Ninomiya fermion doubling on a lattice. It is not clear before the partition function \ur{Z} is better understood whether one needs two `flavors' to reproduce the standard gravity in the infrared region, or one `flavor' is enough.

Nevertheless, one can ask a question in what dimensions $d$ the number of dof's in the frame field (being $d^2$) matches exactly the number of dof's in the spinor representations of the $SO(d)$ Lorentz group. For even $d$, the dimension of the spinor
representation is $d_f=2^{\frac{d}{2}-1}$ and there are two spinor representations, so we double this number. For ``leap'' dimensions, $d=4n$, these two spinor representations are real whereas for $d=4n+2$ they are complex, see {\it e.g.}~\cite{Zee:2003}. Equating
\beq
d^2 = 2^{\frac{d}{2}}\cdot \left\{\begin{array}{ccc} 1,\quad d&=&4n \\ 2,\quad d&=&4n+2 \end{array}\right.
\la{dofs}\eeq
we find that there are only two solutions: $d=2$ and $d=16$. In these dimensions, the number of dof's in the spinor representations is exactly equal to the number of dof's in the frame field, meaning that the composite metric tensor will be unrestricted, and there will be no extra dof's except those needed for gravity. The Cartan formulation is implied here, with the connection $\omega_\mu$ being the independent gauge field of the $SO(d)$ group.

Lattice regularization of gravity suggested in Section V is easily generalizable to any number of dimensions, in particular to $d=16$. The only difference is that now one can
add to the cosmological term $(e\wedge e \wedge\ldots\wedge e)$ (16 factors of $e$) and the Einstein--Hilbert--Cartan term $({\cal F}\wedge e \wedge\ldots \wedge e)$ (14 factors of $e$) six more terms on the same footing, all the way till $({\cal F}\wedge {\cal F} \wedge \ldots \wedge {\cal F}\wedge e\wedge e)$ with arbitrary dimensionless coefficients.
The last term in the sequence, $({\cal F}\wedge\ldots\wedge{\cal F})$, is a full derivative and can be dropped.

Such an action may have various classical saddle points, {\it e.g.} a direct product of spheres or whatever. Similar to the classical Higgs field, such solutions will break the rotational $SO(16)$ symmetry down to smaller gauge groups, for example to $SO(4)_{\rm Lorentz}\times SU(3)_{\rm color}\times SU(2)_{\rm weak}\times U(1)_{\rm Y}$. It should be noted that the 256 dof's of two real 128-dimensional spinor representations of $SO(16)$ fit precisely four generations of the Standard Model.

\section{Conclusions}

Inclusion of fermions into General Relativity requires the introduction of the frame field
$e^A_\mu$ and of the spin connection $\omega^{AB}_\mu$ as independent variables. This is known for the last 90 years as Cartan's formulation of General Relativity.

In Cartan formulation, none of the thinkable local action terms preserving general covariance is sign-definite. It makes impossible to define a quantum gravity theory allowing arbitrarily large fluctuations of metrics and connection about their classical values. 

I call it the Sign Problem of quantum gravity and it seems to be its main difficulty. However, it can be overcome if the frame field $e^A_\mu$ is replaced by a composite field built as a bilinear combination of anticommuting spinor fields and a covariant derivative. The last circumstance is important to ensure that $e^A_\mu$ transforms homogeneously under local Lorentz transformation, as it should do according to the Dirac--Fock--Weyl action for fermions.

One can then formulate a well-defined path integral for the Euclidian partition function
of quantum gravity, regularizing it by a imposing a lattice grid in curved space. The Sign Problem is solved since the Berezin integral over anticommuting spinor fields is well-behaved and finite for any sign of the action. Lattice regularization preserves
exactly local Lorentz gauge invariance and restores diffeomorphism-invariance in the
continuum limit. 

The lattice-regularized partition function with multi-fermion vertices but no propagator defines an interesting new system with rich dynamics that needs to be investigated from all angles. A preliminary study shows that it is exactly equivalent to a sum over closed loops. Whether this system that is well-behaved in the ultraviolet, has the needed classical Einstein's General Relativity in the infrared limit, is a question if the closed loops are long enough in lattice units, to ensure the expected long-range correlations.

Finally, we speculate that spinor fields used to construct the frame field are in fact
matter fermions. By matching the degrees of freedom we find that a regularized diffeomorphism-invariant theory in 16 dimensions with the SO(16) gauge Lorentz group is privileged among other dimensions. The two 128-dimensional spinor representations of SO(16) fit exactly into four generations of fermions. The role of the Higgs phenomenon may be played by mean fields that break the rotational $SO(16)$ symmetry down to the $SO(4)$ Lorentz group and presumably to the $SU(3)\times SU(2)\times U(1)$ gauge sector of the Standard Model.

\vskip 1.5true cm

\noindent
{\bf Acknowledgments.} I thank Victor Petrov, Alexander Tumanov and Alexey Vladimirov
for their help in this project. This work has been presented at the Euler Symposium
in Theoretical and Mathematical Physics (St. Petersburg, July 8 -- 13, 2011) and at the
workshop ``Strings, Gauge Theory and the LHC'' (Copenhagen, Aug. 22 -- Sep. 2, 2011).
I am grateful to the participants of the two meetings for helpful discussions, especially to Herbert Neuberger, Holger Nielsen, Peter Orland, Gerard 't Hooft and Grigory Volovik. This work has been supported in part by Russian Government grants RFBR-09-02-01198 and RSGSS-3628.2008.2, and by Deutsche Forschungsgemeinschaft (DFG) grant 436 RUS 113/998/01.

%\setcounter{section}{0}
%\appendix
%\section{}
% http://www.pdmi.ras.ru/EIMI/2011/STMP/presentations/Diakonov.ppt
% %https://indico.nbi.ku.dk/getFile.py/access?contribId=8&resId=0&materialId=slides&confId=295


\begin{thebibliography}{99}

%\cite{Cartan:1923}
\bibitem{Cartan:1923}
E.~Cartan,
{\it Sur les vari\'et\'es \`a connexion affine, et la th\'eorie de la relativit\`e
g\'en\'eralis\'ee (premi\`ere partie)},
Ann. Sci. de l'\'Ecole Normale Sup\'erieure {\bf 40} (1923) 325 – 412
[English translation by A. Magnon and A. Ashtekar, Bibliopolis, Napoli (1986)].

%\cite{Diakonov:2011fs}
\bibitem{Diakonov:2011fs}
  D.~Diakonov, A.~G.~Tumanov and A.~A.~Vladimirov,
  %``Low-energy General Relativity with torsion: A Systematic derivative
  %expansion,''
  arXiv:1104.2432 [hep-th].
  %%CITATION = ARXIV:1104.2432;%%

%\cite{Baekler:2011jt}
\bibitem{Baekler:2011jt}
  P.~Baekler and F.~W.~Hehl,
  %``Beyond Einstein-Cartan gravity: Quadratic torsion and curvature invariants
  %with even and odd parity including all boundary terms,''
  arXiv:1105.3504 [gr-qc].
  %%CITATION = ARXIV:1105.3504;%%

\bibitem{Berezin:1966}
F.A.~Berezin, {\it The Method of Second Quantization}, Academic Press, New York (1966).

%\cite{Akama:1978pg}
\bibitem{Akama:1978pg}
  K.~Akama,
  %``AN ATTEMPT AT PREGEOMETRY: GRAVITY WITH COMPOSITE METRIC,''
  Prog.\ Theor.\ Phys.\  {\bf 60} (1978) 1900.
  %%CITATION = PTPKA,60,1900;%%

\bibitem{Volovik:1990}
G.E.~Volovik, Physica {\bf B162} (1990) 222.

%\cite{Wetterich:2005yi}
\bibitem{Wetterich:2005yi}
  C.~Wetterich,
  %``Spontaneous symmetry breaking origin for the difference between time and
  %space,''
  Phys.\ Rev.\ Lett.\  {\bf 94} (2005) 011602.
  %%CITATION = PRLTA,94,011602;%%

%\cite{Wetterich:2011yf}
\bibitem{Wetterich:2011yf}
  C.~Wetterich,
  %``Lattice spinor gravity,''
  arXiv:1108.1313 [hep-th].
  %%CITATION = ARXIV:1108.1313;%%

%\cite{Fock:1929vt}
\bibitem{Fock:1929vt}
V.~Fock and D.~Iwanenko,
% ``Linear Quantum Geometry and Parallel Transport,''
  C.\ R.\ Acad.\ Sci., Paris {\bf 188} (1929) 1470;
V.~Fock,
  % ``Geometrization of Dirac's theory of the electron,''
  Z.\ Phys.\ {\bf 57} (1929) 261;
  %%CITATION = ZEPYA,57,261;%%
V.~Fock,
  % ``On the Dirac Equations in General Relativity,''
  C.\ R.\ Acad.\ Sci., Paris {\bf 189} (1929) 25;
V.~Fock,
% ``Dirac Wave Equation and Riemann Geometry'',
Le Journal de Physique et de Radium, S\`erie VI, {\bf 10} (1929) 392.
[English translation of the last two papers in: V.A.~Fock, Selected works,
L.D.~Faddeev, L.A~Khalfin and I.V.~Komarov, eds., Chapman and Hall
% / CRC
(2004)].

%\cite{Weyl:1929fm}
\bibitem{Weyl:1929fm}
  H.~Weyl,
  % ``Electron and Gravitation I (In German),''
  Z.\ Phys.\  {\bf 56} (1929) 330
  [Surveys High Energ.\ Phys.\  {\bf 5} (1986) 261].
  %%CITATION = SHEPD,5,261;%%

%\cite{Hojman:1980kv} Nelson:1980ph
\bibitem{Hojman:1980kv}
  R.~Hojman, C.~Mukku and W.~A.~Sayed,
  %``PARITY VIOLATION IN METRIC TORSION THEORIES OF GRAVITATION,''
  Phys.\ Rev.\  D {\bf 22} (1980) 1915.
  %%CITATION = PHRVA,D22,1915;%%

%\cite{Nelson:1980ph}
\bibitem{Nelson:1980ph}
  P.~C.~Nelson,
  %``GRAVITY WITH PROPAGATING PSEUDOSCALAR TORSION,''
  Phys.\ Lett.\  A {\bf 79}, 285 (1980).
  %%CITATION = PHLTA,A79,285;%%

\bibitem{footnote-1}
In Ref.~\cite{Wetterich:2011yf} local Lorentz symmetry is achieved despite the wrong
transformation properties of the frame field, by taking a specific action.

\bibitem{Zee:2003}
A.~Zee, {\it Quantum Field Theory in a Nutshell}, Princeton University Press (2003), Ch. VII.

\end{thebibliography}
\end{document}